\documentclass[12pt]{article}
\usepackage{amsmath,leftidx}
\usepackage{amssymb}
\usepackage{amsthm}
\usepackage{fullpage}
\usepackage[utf8]{inputenc}
\usepackage{mathtools}
\usepackage{url}
\usepackage[usenames,svgnames]{xcolor}
\usepackage{cite}
\usepackage{enumitem}
\usepackage[titletoc,title]{appendix}
\usepackage{comment}
\usepackage{mathdots}
\usepackage{graphicx}

\usepackage[margin=1.3in]{geometry}

\usepackage{ tikz, pgfplots, todonotes,color}
\usetikzlibrary{decorations.markings,arrows,}

\usepackage{mathtools,cancel}
\usepackage{url}

\usepackage{todonotes}
\usepackage{comment}
\usepackage{mathdots}
\usepackage{graphicx}
\usepackage{float}
\usepackage{enumitem}

\usepackage[pagebackref=false]{hyperref} 
\usepackage[all]{hypcap} 

\theoremstyle{plain}
\newtheorem{theorem}{Theorem}[section]

\newtheorem{definition-theorem}[theorem]{Definition-Theorem}
\newtheorem{definition-proposition}[theorem]{Definition-Proposition}

\newtheorem{corollary}[theorem]{Corollary}
\newtheorem{example}{Example}[section]
\newtheorem{examples}{Example}[subsection]

\newtheorem{remark}{Remark}[section]

\theoremstyle{definition}

\numberwithin{equation}{section} 

\DeclareMathOperator{\tr}{tr}

\DeclareMathOperator{\cyc}{cyc}

\DeclareMathOperator{\aut}{aut}

\def\ra{{\rightarrow}}

\def\tr{\mathrm {tr}}
\def\det{\mathrm {det}}

\def\Mat{\mathrm {Mat}}

\def\Nor{\mathrm {Nor}}

\def\diag{\mathrm {diag}}

\def\be{\begin{equation}}
\def\ee{\end{equation}}

\def\bea{\begin{eqnarray}}
\def\eea{\end{eqnarray}}

\def\bt{\begin{theorem}}
\def\et{\end{theorem}}

\def\bex{\begin{example}\small \rm}
\def\eex{\end{example}}

\def\bexs{\begin{examples}\small \rm}
\def\eexs{\end{examples}}

\def\ra{\rightarrow}

\def\br{\begin{remark}\small \rm}
\def\er{\end{remark}}

\def\&{&{\hskip -20pt}}

\def\CC{\mathcal{C}}
\def\DD{\mathcal{D}}

\def\LL{\mathcal{L}}

\def\WW{\mathcal{W}}

\def\Cb{\mathbf{C}}

\def\Nb{\mathbf{N}}

\def\Nb{\mathbf{N}}
\def\Pb{\mathbf{P}}

\def\Zb{\mathbf{Z}}

\begin{document}
\baselineskip 16pt

\medskip
\begin{center}
\begin{Large}\fontfamily{cmss}
\fontsize{17pt}{22pt}
\selectfont
\textbf{Weighted Hurwitz numbers, $\tau$-functions \\ and matrix integrals$^{\dag}$\footnote {Text of invited
presentation  at: Quantum Theory and Symmetries,  XIth International symposium,  Centre de recherches math\'ematiques, Montr\'eal,  July 1-5, 2019.}}
	\end{Large}
\\
\bigskip \bigskip
\begin{large}  J. Harnad$^{1, 2}$\footnote {e-mail: harnad@crm.umontreal.ca  }
 \end{large}\\
\bigskip
\begin{small}
$^{1}${\em Department of Mathematics and Statistics, Concordia University\\ 1455 de Maisonneuve Blvd.~W.~Montreal, QC H3G 1M8  Canada}\\
\smallskip
$^{2}${\em Centre de recherches math\'ematiques, Universit\'e de Montr\'eal, \\C.~P.~6128, succ. centre ville, Montr\'eal, QC H3C 3J7  Canada}\\
\smallskip
\end{small}
\end{center}

\begin{small}\begin{center}
\end{center}\end{small}

\begin{abstract}
 The basis elements spanning the  Sato Grassmannian element corresponding
to the KP $\tau$-function that serves as generating function for rationally weighted Hurwitz numbers are shown to be Meijer $G$-functions.
Using their Mellin-Barnes integral representation the $\tau$-function, evaluated at the  trace invariants of an externally coupled matrix, 
is expressed as  a matrix integral.  Using the Mellin-Barnes integral transform of an infinite product of $\Gamma$ functions,
a similar matrix integral representation is given for the KP $\tau$-function  that serves as
generating function for quantum weighted Hurwitz numbers. \end{abstract}

 \section{Hurwitz numbers: classical and weighted}
 
The fact that KP and $2D$-Toda $\tau$-functions of hypergeometric type serve as generating functions 
for {\em weighted Hurwitz numbers} was shown in   \cite{GH1, HO1, H1, GH2}, generalizing the case of {\em simple }
(single and double) Hurwitz numbers \cite{Ok, OP}.
 Sections \ref{geometric_hurwitz} and \ref{weighted_hurwitz} below, and Section \ref{hypergeom_gen_fns}  
 give a brief review of this theory, together with two illustrative examples: rational and quantum weighted Hurwitz numbers.
 In Section \ref{wronskian_matrix_int}, it is shown how evaluation of such $\tau$-functions at the trace invariants
 of a finite matrix may be expressed either as a Wronskian determinant or as a matrix integral.
 The content of subsections \ref{quant_classical_curve}--\ref{matrix_int_rep} are largely drawn from \cite{BH, HR}, 
 in which further details and proofs of the main results may be found.
 

 \subsection{Geometric meaning of classical Hurwitz numbers}
 \label{geometric_hurwitz}
 
 The Hurwitz number  $H(\mu^{(1)}, \dots , \mu^{(k)})$  is the number of inequivalent branched $N$-sheeted
covers $\Gamma \ra \Pb^1$ of the Riemann sphere, with $k$ branch points $(Q_1, \dots, Q_k)$, 
whose {\em ramification profiles } are given by $k$ {\em partitions} $(\mu^{(1)}, \dots , \mu^{(k)})$ of $N$,
normalized by dividing by the order $|\aut(\Gamma)|$ of its automorphism group.
The {\em Euler characteristic} $\chi$ and {\em genus} $g$ of the covering curve is given 
by the {\em  Riemann-Hurwitz formula}:
\be
\chi =2-2g = 2N -  d, \quad  d:= \sum_{i=1}^l \ell^*(\mu^{(i)}), 
\ee
 where  $\ell^*(\mu) := | \mu |  -  \ell(\mu) =N - \ell(\mu) $  is the {\em colength} of the partition.

The {\em Frobenius-Schur} formula gives $H(\mu^{(1)}, \dots \mu^{(k)})$ in terms of $S_N$ characters:
\be
H(\mu^{(1)}, \dots \mu^{(k)}) =  \sum_{\lambda, |\lambda|= N} h^{k-2}(\lambda)
 \prod_{j=1}^k{\chi_\lambda(\mu^{(i)})\over z_{\mu^{(j)}}} ,  \quad |\mu^{(i)}| = N , 
\ee
where
$h(\lambda) = \left(\det {1 \over (\lambda_i - i +j)!}\right)^{-1}$
is the  product of the hook lengths of the partition 
$\lambda = (\lambda_1 \ge \cdots \ge \lambda_{\ell(\lambda} >0)$,
\quad $\chi_\lambda(\mu^{(j)})$
is the {\em irreducible character} of representation $\lambda$ evaluated on the conjugacy class $\mu^{(j)}$, and
\be
z_{\mu^{(j)}} := \prod_{i} i^{m_i(\mu^{(j)})}( m_i(\mu^{(j)}))! 
\ee
is the order of the stabilizer of any element of $\cyc(\mu^{(j)})$  
(and $m_i(\mu^{(j)})=$ \# parts  of partition $\mu^{(j)}$  equal to $ i$)
    
\subsection{Weighted Hurwitz numbers \cite{GH1, HO1, H1, GH2}}
\label{weighted_hurwitz}

Define the  {\em weight generating function} $G(z)$, or its dual $\widetilde{G}(z)$,
as an infinite (or finite)  product or sum (formal or convergent).
\bea
G(z) &\&= \prod_{i=1}^\infty(1+ z c_i) = 1 +\sum_{j=1}^\infty g_j z^j \cr
  \widetilde{G}(z) &\&= \prod_{i=1}^\infty(1- z c_i)^{-1}    = 1 +\sum_{j=1}^\infty \widetilde{g}_j z^j.
\eea
The  weight for a branched covering  with ramification profiles 
$(\mu^{(1)}, \dots, \mu^{(k)})$  is defined to be:
\bea
\WW_G(\mu^{(1)}, \dots, \mu^{(k)}) &\& :=
{1\over k!} \sum_{\sigma\in S_k} \sum_{1 \le i_1 < \cdots < i_k}
 c_{i_{\sigma(1)}}^{\ell^*(\mu^{(1)})} \cdots c_{i_{\sigma(k)}}^{\ell^*(\mu^{(k)})} \cr
\widetilde{\WW}_{\widetilde{G}}(\mu^{(1)}, \dots, \mu^{(k)}) &\& :=
{(-1)^{\sum_{i=1}^k \ell^*(\mu^{(i)})+k}\over k!} \sum_{\sigma\in S_k} \sum_{1 \le i_1 \le \cdots \le i_k}
 c_{i_{\sigma(1)}}^{\ell^*(\mu^{(1)})} \cdots c_{i_{\sigma(k)}}^{\ell^*(\mu^{(k)})}
\eea
    
 {\em Weighted double Hurwitz numbers} $H^d_G(\mu, \nu)$, $H^d_{\widetilde{G}}(\mu, \nu)$
 for $n$-sheeted branched  coverings of the Riemann sphere  having a 
  pair of unweighted branch points $(Q_0, Q_\infty)$, with ramification profiles of type 
$(\mu, \nu)$, and  $k$  additional weighted branch points $(Q_1, \dots, Q_k)$
with ramification profiles $(\mu^{(1)}, \dots, \mu^{(k)})$ are defined as:
\bea
H^d_G(\mu, \nu) &\&:= \sum_{k=1}^d \sideset{}{'}\sum_{\substack{\mu^{(1)}, \dots \mu^{(k)} \\ \sum_{i=1}^k \ell^*(\mu^{(i)})= d}}
\WW_{G}(\mu^{(1)}, \dots, \mu^{(k)}) H(\mu^{(1)}, \dots, \mu^{(k)}, \mu, \nu) ,  \cr
H^d_{\widetilde{G}}(\mu, \nu) &\&:= \sum_{k=1}^d \sideset{}{'}\sum_{\substack{\mu^{(1)}, \dots \mu^{(k)} \\ \sum_{i=1}^k \ell^*(\mu^{(i)})= d}}
\widetilde{\WW}_{\widetilde{G}}(\mu^{(1)}, \dots, \mu^{(k)}) H(\mu^{(1)}, \dots, \mu^{(k)}, \mu, \nu) ,
\nonumber
\eea
where $\sum'$ denotes the sum over all partitions  other than the cycle type of the identity element $(1)^n$.
     If $Q_\infty$ is not a branch point; i.e. $\nu = (1)^n$,  we have a weighted  {\em single} Hurwitz number
     \be
     H^d_G(\mu) := H^d_G(\mu, (1)^n).
     \ee

Two cases of particular interest are:   
            {\em rational weight generating functions:}
        \be 
        G_{{\bf c}, {\bf d}}(z) :={ \prod_{l=1}^L (1+c_l z) \over \prod_{m=1}^M(1-d_m z)}.
        \ee
     and  {\em quantum weight generating function (quantum exponential):}
\be
\widetilde{G}(z) = H_q(z) := \prod_{i=0}^\infty(1 - q^i z)^{-1} = \sum_{n=0}^\infty {z^n \over (q; q)_n},
\ee
where
\be
(q ; q)_n := (1-q)(1-q^2) \cdots (1-q^n)
\ee
for some parameter $q$, with $|q|<1$.
           
The corresponding {\em rationally weighted} (single) Hurwitz numbers are:
       \bea
H^d_{G_{{\bf c}, {\bf d}}}(\mu, \nu)&\& := 
\sum_{1\le k, l \atop k+l \le d} \sum_{{\mu^{(1)}, \dots \mu^{(k)} ,\nu^{(1)}, \dots \nu^{(l)}, \atop    
\sum_{i=1}^k \ell^*(\mu^{(i)})+ \sum_{j=1}^l \ell^*(\nu^{(j)})  =d}\atop  |\mu^{(i)}|  =   |\nu^{(j)}|=N }
{\hskip -30 pt} \WW_{G_{{\bf c}, {\bf d}}}(\mu^{(1)}, \dots, \mu^{(k)}; \nu^{(1)}, \dots, \nu^{(l)})  \cr
&\&{\hskip 120 pt}  \times H(\mu^{(1)}, \dots, \mu^{(k)}, \nu^{(1)}, \dots, \nu^{(l)}, \mu),
\nonumber
\eea        
where the rational weight factor is:
      \bea
&\& \WW_{G_{{\bf c}, {\bf d}}}(\mu^{(1)}, \dots, \mu^{(k)}; \nu^{(1)}, \dots, \nu^{(l)}) \cr
&\& :=
{(-1)^{\sum_{j=1}^l \ell^*(\nu^{(j)}) -l }\over k! l!}
 \sum_{\sigma \in S_k \atop \sigma' \in S_l}   \sum_{1 \leq a_1 < \cdots < a_{k } \leq M\atop 1 \leq b_1\cdots \leq b_k\leq L } 
  c_{a_{\sigma(1)}}^{\ell^*(\mu^{(1)})} \cdots c_{a_{\sigma(k)}}^{\ell^*(\mu^{(k)})}  
  d_{b_{\sigma'(1)}}^{\ell^*(\nu^{(1)})} \cdots d_{b_{\sigma'(l)}}^{\ell^*(\nu^{(l)})}.
  \nonumber
  \eea
       The    {\em quantum weighted} (single) Hurwitz numbers are
\be
H^d_{H_q}(\mu) := 
\sum_{k=1}^d \sum_{\mu^(1), \dots \mu^{(k)}, \  |\mu^{(i)}| =N  \atop \sum_{i=1}^k \ell^*(\mu^{(i)}) =d} 
\widetilde{\WW}_{H_q}(\mu^{(1)}, \dots, \mu^{(k)}) H(\mu^{(1)}, \dots, \mu^{k)}, \mu),
\ee
 where the    {\em quantum weight factor} is
      \bea
 \widetilde{\WW}_{H_q}(\mu^{(1)}, \dots, \mu^{(k)})&\&:=
{(-1)^{d-k}\over k!} \sum_{\sigma \in S_k} \prod_{j=1}^k  {1  \over (1 - q^{\sum_{i=1}^j\ell^*(\mu^{(\sigma(i))}})}.
\nonumber
\eea


  \section{Hypergeometric $\tau$-functions as generating functions for weighted Hurwitz numbers \cite{GH1, HO1, H1, GH2}}
  \label{hypergeom_gen_fns}

To construct a KP  $\tau$-function of hypergeometric type that serves as generating function
for weighted Hurwitz numbers for a given weight generating function $G$, choose a  small parameter $\beta$
and  define coefficients $r^{(G, \beta)}_\lambda$ that are of {\em content product form}:
\be
 r^{(G,\beta)}_\lambda :=  \prod_{(ij)\in \lambda} r_{j-i}^{(G,\beta)}=  \prod_{(ij)\in \lambda} G((j-i)\beta) ,   
 \ee
where 
\be  
r_j^{(G, \beta)}  := G(j\beta) = {\rho^{(G,\beta)}_j \over\beta  \rho^{(G, \beta)}_{j-1}} ,
\ee
with
\bea
\rho^{(G, \beta)}_j &\&:=\beta^j\prod_{i=1}^j G(i\beta) =: e ^{T_j^G(\beta)},  \quad \rho_0=1, 
= {\rho^{(G,\beta)}_j \over\beta  \rho^{(G, \beta)}_{j-1}} ,\cr
 \rho^{(G, \beta)}_{-j}(&\& := \beta^{-j} \prod_{i=1}^{j-1} {1\over G(-i\beta)} =: e ^{T_{-j}^G(\beta)},  \quad j=  1, 2, \dots 
 \eea
 We then have \cite{GH2, HO1}:
   
\begin{theorem}[Hypergeometric  Toda $\tau$-functions associated to weight generating function $G(z)$]

The double Schur function series
\be
\tau^{( G, \beta)}({\bf t}, {\bf s}) :=   \sum_{\lambda }\beta^{|\lambda|}r_\lambda^{(G,\beta)} s_\lambda({\bf t}) s_\lambda({\bf s}) \
\ee
defines a $2D$-Toda $\tau$-function (at lattice value $n=0$) .
 \end{theorem}
 
     We now use  the {\em Frobenius character formula}
 \be
s_\lambda({\bf t}) = \sum_{\mu, |\mu|=|\lambda |}{ \chi_\lambda(\mu) p_\mu({\bf t}) \over z_\mu}, \quad
s_\lambda({\bf s}) = \sum_{\nu, |\nu|=|\lambda |}{ \chi_\lambda(\nu) p_\nu({\bf s}) \over z_\nu}
\ee
 to change the basis of Schur functions  to {\em power sum symmetric functions} 
\be
p_\mu({\bf t}) := \prod_{i=1}^{\ell(\mu)} p_{\mu_i}({\bf t}), \ p_j ({\bf t})= j t_j,
\quad p_\nu({\bf s}) := \prod_{i=1}^{\ell(\nu)} p_{\nu_i}({\bf s}), \ p_j ({\bf s})= j s_j.
\ee
  
\begin{theorem}[Hypergeometric Toda $\tau$-functions as generating function for weighted double Hurwitz numbers \cite{GH2, HO1}]
The $\tau$-function $\tau^{( G, \beta)}({\bf t}, {\bf s})$ can  equivalently be
expressed as a double infinite series in the bases of  power sum symmetric functions  as follows
\be
\tau^{( G, \beta)}({\bf t}, {\bf s}) =  \sum_{d=0}^\infty \sum_{\mu, \nu, \atop |\mu|=|\nu|}\beta^{|\mu|+d} H_G^d(\mu, \nu) p_\mu({\bf t}) p_\nu({\bf s}) .
\ee
It is thus a generating function for the numbers $H_G^d(\mu, \nu) $ of weighted $ n$-fold branched coverings
of the sphere, with a pair of specified branch points having ramification profiles $(\mu, \nu)$ and genus given
by the Riemann-Hurwitz formula 
\be
2-2g = \ell(\mu) + \ell(\nu) - d,  \quad d =\sum_{i=1}^k\ell^*(\mu^{(i)}).
\ee
 \end{theorem}

\begin{corollary}[Hypergeometric KP $\tau$-functions as generating functions for weighted single Hurwitz numbers]
Set: \quad  ${\bf s} = \beta^{-1} {\bf t}_0 := (\beta^{-1}, 0, 0, \dots)$. \hfil
\break \noindent Then the series
\bea
\tau^{( G, \beta)}({\bf t}, \beta^{-1}{\bf t}_0) &\&:= \tau^{( G, \beta)}({\bf t}) = \sum_{\lambda }(h(\lambda))^{-1}r_\lambda^{(G, \beta)} s_\lambda({\bf t})  \cr
&\&=  \sum_{d=0}^\infty \sum_{\mu}\beta^d H_G^d(\mu) p_\mu({\bf t})
\nonumber
\eea
is a $KP$ $\tau$-function which is a generating function for weighted single  numbers $H_G^d(\mu) $ for $|\mu|$-fold branched coverings
of the sphere, with a branch point having ramification profile $(\mu)$ at $Q_0$ and genus given
by the Riemann-Hurwitz formula.
\be
2-2g = |\mu|+ \ell(\mu) - d.
\ee
 \end{corollary}
 
\section{Wronskian and matrix integral representation of $\tau^{(G,\beta)}([X])$}
\label{wronskian_matrix_int}

In \cite{BH, HR} new matrix integral representations  were derived for the $\tau$-functions that serve as generating
functions for rationally and quantum weighted Hurwitz numbers.  The main result is that, using
Laurent series and Mellin-Barnes integral representations of the adapted bases for the respective elements of the infinite Grassmannian
corresponding to these cases, the $\tau$-functions may be expressed as Wronskian determinants or as matrix integrals.

 \subsection{Adapted basis, recursion operators, quantum spectral curve}
 \label{adaapted_basis_rec_op}
 
Henceforth, we always set: 
\be
{\bf s} = \beta^{-1} {\bf t}_0 := (\beta^{-1}, 0, 0, \dots)
\ee
and
\be
\tau^{(G, \beta)}({\bf t}) := \tau^{(G, \beta)}({\bf t}, \beta^{-1}{\bf t}_0)
\ee
 is a KP $\tau$-function of hypergeometric type.
 
  For  $k \in \Zb$, define:
\bea
\phi_k(x) &\&:= {\beta\over 2\pi i  x^{k-1}} \oint_{|\zeta|=\epsilon} \rho^{(G, \beta)}(\zeta) e^{\beta^{-1} x\zeta } {d\zeta\over \zeta^k}, \cr
&\& = \beta x^{1-k}\sum_{j= 0}^{\infty} { \rho^{(G,\beta)}_{j-k}\over j!} \left({x \over\beta}\right)^j,
\eea
where
\be
\rho^{(G, \beta)}(\zeta) := \sum_{i=-\infty}^{k-1} \rho^{(G, \beta)}_{-i-1} \zeta^i.
\ee
Then $\{\phi_k(1/z)\}_{k\in \Nb^+}$ is a basis for the element $w^{(G,\beta)}$ of the Sato Grassmannian
that determines the KP $\tau$-function $\tau^{(G, \beta)}({\bf t})$ \cite{ACEH1}.   
   
 
 \subsection{Quantum and classical spectral curve}
 \label{quant_classical_curve}
   
\begin{theorem}[Quantum spectral curve and eigenvalue equations \cite{ACEH1}]
The functions $\phi_k(x)$  satisfy 
\be
\LL \phi_k(x):= \left( x G(\beta\DD) - \DD \right)\phi_k(x)=(k-1)\phi_k(x)
\ee
where $\DD := x{d\over dx}$ is the Euler operator.
\end{theorem}
The classical spectral curve is
\be
y = G(\beta xy).
\ee

   \subsubsection {Rational weighting case}
For $G(z) = G_{{\bf c}, {\bf d}}(z)$, \ denote $\phi_k(x) =: \phi_k^{({\bf c}, {\bf d},\beta)}(x)$. Then
\be
  \zeta \prod_{l=1}^L (\DD + {1\over \beta c_l })\phi^{({\bf c},{\bf d},\beta)}_k
  + (\DD+k-1)  \prod_{m=1}^M(\DD  -1 -{1\over \beta d_m} )\phi^{({\bf c},{\bf d},\beta)}_k =0,
\label{phi_k_eq_dir_zeta_meijer}
\ee
where
\be
\zeta :=  -\kappa_{{\bf c}, {\bf d}}x, \quad \kappa_{{\bf c}, {\bf d}}:=(-1)^{M} {\prod_{l=1}^L \beta c_l\over \prod_{m=1}^M \beta d_m}.
\ee

\subsubsection{Mellin-Barnes integral repesentation: Meijer $G$-functions \cite{BH, HR}}
\begin{figure}[H]
\label{fig1}
\begin{center}
\resizebox{0.6\textwidth}{!}{
\begin{tikzpicture}[scale=01.5]

\draw[fill] (0,0) node[below]{$-k+1$} circle [radius=0.03];
\draw[fill] (1,0) node[below]{$-k+2$} circle [radius=0.03];
\draw[fill] (2,0) node[below]{$\cdots$} circle [radius=0.03];
\draw[fill] (3,0) node[below]{$\cdots $} circle [radius=0.03];
\draw[fill] (4,0) node[below]{$N$} circle [radius=0.03];
\draw[fill] (5,0) node[below]{$N+1$} circle [radius=0.03];

\draw[fill,red] (1.3,1) node[below]{$-\frac 1{\beta c_\ell}$} circle [radius=0.03];
\draw[fill,red] (0.3,1) node[below]{$-\frac 1{\beta c_\ell}-1$} circle [radius=0.03];
\draw[fill,red] (-0.7,1) node[below]{$\cdots$} circle [radius=0.03];
\draw[fill,red] (-1.7,1) node[below]{$\cdots $} circle [radius=0.03];

\draw[fill,red] (0.7,-0.7) node[below]{$-\frac 1{\beta c_\ell}$} circle [radius=0.03];
\draw[fill,red] (-0.4,-0.7) node[below]{$-\frac 1{\beta c_\ell}-1$} circle [radius=0.03];
\draw[fill,red] (-1.3,-0.7) node[below]{$\cdots$} circle [radius=0.03];
\draw[fill,red] (-2.3,-0.7) node[below]{$\cdots $} circle [radius=0.03];

\draw[->] (2,-3)--(2,3);
\draw[->] (-3,0)--(5.5,0);

\draw[dashed, -> ] (4.5,-3)--node[pos =0.2,sloped, above] {$\Re s= N+\frac 1 2$} (4.5,3);

\draw[ blue,  line width=1,postaction={decorate,decoration={{markings,mark=at position 0.75 with {\arrow[black,line width=1.5pt]{>}}}} }]
(4.5,-3)--node[pos=0.6, below, sloped]{$\Re s = N+\frac 1 2 $}(4.5, 3);

\draw[black , line width =1,postaction={decorate,decoration={{markings,mark=at position 0.75 with {\arrow[black,line width=1.5pt]{>}}}} }]
(4.49,-3) to (4.49, -1) to[out=90, in = -90, looseness=0.5] (-0.3,0)
to [out=90, in =-90, looseness=0.5] node [pos=0.7, sloped, above]{ $\mathcal C_k$} (4.49, 1) to (4.49,3);
;

\end{tikzpicture}}
\end{center}
\caption{The contours of integration for the function $\phi_k^{({{\bf c}, {\bf d},\beta})}$ in the case $L>M+1$.}
\label{Ck}
\end{figure}
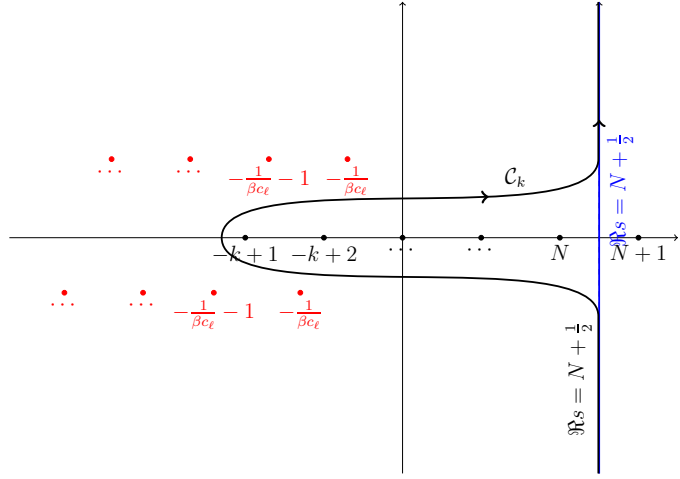

It may be shown that $\phi_k^{({{\bf c}, {\bf d},\beta})} $ has the Mellin-Barnes integral representation:
\bea
\phi_k^{({{\bf c}, {\bf d},\beta})} &\&
=  C_k^{({\bf c},{\bf d},\beta)} G_{L,M+1}^{1,L}\left( {1-\frac 1{\beta c_1}, \cdots, 1-\frac 1{\beta c_L} \atop 1-k,
1+\frac 1 {\beta d_1}, \cdots, 1+\frac 1{\beta d_M}}  \bigg{|}  -\kappa_{{\bf c}, {\bf d}} x \right) \cr
&\&={C_k^{({\bf c},{\bf d},\beta)}\over 2\pi i } \int_{\CC_k}  \frac {\Gamma(1-k-s)  \prod_{\ell=1}^L \Gamma \left( s + \frac 1 {\beta c_\ell} \right) 
\left(-\kappa_{{\bf c}, {\bf d}} x\right)^s}{\prod_{m=1}^M \Gamma\left( s - \frac 1{\beta d_m}  \right)}  ds. \cr
&\& \sim {\beta \rho_{-k}({\bf c }, {\bf d})\over( \kappa x)^{k-1}}
\leftidx{_L}{F_M}\left( {1-k+{1 \over \beta c_1}, \cdots, 1-k+{1\over \beta c_L} \atop
1-k-{1 \over \beta d_1}, \cdots,1-k- {1 \over \beta d_M}}  \bigg{|}  \kappa_{{\bf c}, {\bf d}} x \right)
\eea
where
\be
C_k^{({\bf c},{\bf d},\beta)} :={\prod_{j=1}^M \Gamma(-{1\over \beta d_j}) \over (-\beta)^{k-1}\prod_{\ell=1}^L \Gamma({1\over \beta c_\ell})},
\ee
The contour $\CC_k$ is chosen so that the  poles at $1-k, 2-k, \cdots$ are to the right
and the poles at $\{-i- {1\over \beta c_j}\}_{ j=1, \cdots L, \, i\in \Nb^+}$ to the left. (See Figure \ref{fig1}.)

   \subsubsection  {Quantum case expressed as Mellin-Barnes integrals \cite{HR}}
The following is an integral representation of  $\phi^{(H_q, \beta)}_k(x)$, valid for all $x\in \Cb$,
\be
\phi^{(H_q, \beta)}_k= 
\frac{1}{2\pi i}\int_{\CC_k} A_{H_q, k}(s) x^s ds, 
\ee
where
\be
A_{H_q, k}(z) := (-\beta)^{1-k}\Gamma(1-k-z)\prod_{m=0}^\infty \left( (-\beta q^{m})^{-z} {\Gamma(-\beta^{-1}q^{-m}) \over \Gamma(z-\beta^{-1}q^{-m})}\right).
\ee
The contour $\CC_k$ is defined as starting  at $+\infty$ immediately above the real axis, proceeding to the left above the axis,
winding around the poles at the integers $s= -k, -k+1. \dots$ in a conterclockwise sense and continuing  below the axis back
to $+\infty$.

\subsection{Determinantal representation of $\tau^{(G,\beta)}({\bf t})$}
\label{det_rep}
  
If $\tau^{(G,\beta)}({\bf t})$  is evaluated at the trace invariants of diagonal $X  \in \Mat^{n\times n}$
\bea
{\bf t}&\&= \big[ X \big], \quad t_i = {1\over i} \tr X^i, \cr
X &\&:= \diag(x_1, \dots, x_n), 
\eea
it is expressible as the ratio of $n \times n$ determinants 
\be
\tau^{(G,\beta)}\left(\big[X \big]\right) ={\prod_{i=1}^n x_i^{n-1}\over \prod_{i=1}^n \rho_{-i}} {\det\left( \phi_i(x_j)\right)_{1\leq i,j, \leq n} \over \Delta(x)},
\label{tau_phi_i_det}
\ee
where
\be
\Delta(x) = \prod_{1\leq i < j \leq n}(x_i - x_j) = \det(x_i^{n-j})_{1\leq i,j, \leq n} 
\ee
is the Vandermonde determinant.

   \subsubsection{Eulerian Wronskian representation}
It follows from the recursion relations
\be
\beta (\DD +k -1) \phi_k= \phi_{k-1}, \quad k \in \Zb,
\ee
that
\be
\tau^{(G,\beta)}\left(\big[X \big]\right) =\gamma_n \left(\prod_{i=1}^n x_i^{n-1}\right){\det\left( \DD^{i-1}\phi_n(x_j)\right)_{1\leq i,j, \leq n} \over \Delta(x)},
\ee
where
\be
\gamma_n :={\beta^{{1\over 2}n(n-1)}\over \prod_{i=1}^n \rho_{-i}}.
\ee

\subsection{Matrix integral representation of $\tau^{(G,\beta)}([X])$ \cite{BH, HR}}
\label{matrix_int_rep}

  
    \subsubsection{Wronskian representation: rational case}
For  rational weight generating functions $G=G_{{\bf c}, {\bf d}}$,  and any  $n\in \Nb^+$, let
\bea
\phi^{({\bf c}, {\bf d}, \beta)}_n(e^y) &\&= \int_{\CC_n} A^{({\bf c}, {\bf d}, \beta)}_n(s) e^{ys} ds,
\cr
A^{({\bf c}, {\bf d}, \beta)}_n(s) &\&:= 
{ C_n^{({\bf c},{\bf d},\beta)}\Gamma(1-n-s) \prod_{l=1}^L \Gamma \left(s + {1 \over \beta c_l }\right) (-\kappa_{{\bf c}, {\bf d}})^s \over 
 2\pi i  \prod_{m=1}^M \Gamma\left( s- \frac 1{\beta d_m} \right) }.
 \nonumber
\eea
Define the diagonal matrix $Y= \diag (y_1, \dots y_n)$
\be
X = e^Y, \quad Y=\ln(X), \quad x_i = e^{y_i}, \quad i =1 , \dots, n. 
\ee
Then  $\tau^{(G_{{\bf c}, {\bf d}},\beta)}\left(\big[X \big]\right) $ becomes a ratio of Wronskian determinants
\be
\tau^{(G_{{\bf c}, {\bf d}},\beta)}\left(\big[X \big]\right) =\gamma_n  \left(\prod_{i=1}^n x_i^{n-1}\right){\det\left(( \phi_n^{({\bf c}, {\bf d}, \beta)})^{(i-1)}(e^{y_j})\right)_{1\leq i,j, \leq n} \over \Delta(e^y)}.
\ee

  
    \subsubsection{Matrix integral representation of $\tau^{(G,\beta)}([X])$: rational case}
It follows \cite{{BH}} that
\be
\tau^{(G_{{\bf c}, {\bf d}},\beta)} (\big[ X \big])= {\beta^{{1\over 2}n(n-1)} (\prod_{i=1}^n x_i^{n-1})\Delta(\ln(x)) \over (\prod_{i=1}^n i!) 
\Delta(x) } \Zb_{d\mu_{({\bf c}, {\bf d}, \beta, n)} }(X),
\ee
where
\be 
\Zb_{d\mu_{({\bf c}, {\bf d}, \beta, n)}}(X) =\int_{M \in \Nor^{n\times n}_{\CC_n}}d\mu_{({\bf c}, {\bf d}, \beta, n)}(M) e^{\tr Y M}
\ee
and
\be d\mu_{({\bf c}, {\bf d}, \beta, n)} (M):= (\Delta({\bf \zeta})^2 \det(A^{({\bf c}, {\bf d}, \beta)}_n(M)) d\mu_0(U)\prod_{j=1}^n d\zeta_i 
\nonumber
 \ee
 is a conjugation invariant measure on the space of normal matrices  
 \be
M = U Z U^\dag \in \Nor^{n\times n}_{\CC_n}, \quad U \in U(n), \quad Z= \diag(\zeta_1, \dots, \zeta_n)
\ee
with eigenvalues $\zeta_i \in \Cb$ supported on the contour $\CC_n$.


    \subsubsection{Wronskian representation: quantum case}
For  quantum weight generating functions $G=H_q$,  and any  $n\in \Nb^+$, let
\bea
\phi^{(H_q, \beta)}_n(e^y) &\&= \int_{\CC_n} A^{({\bf c}, {\bf d}, \beta)}_n(s) e^{ys} ds,
\cr
A_{H_q, n}(z) &\&:= (-\beta)^{1-n}\Gamma(1-n-z)\prod_{m=0}^\infty \left( (-\beta q^{m})^{-z} {\Gamma(-\beta^{-1}q^{-m}) \over \Gamma(z-\beta^{-1}q^{-m})}\right).
\nonumber
\eea

Define the diagonal matrix $Y= \diag (y_1, \dots y_n)$
\be
X = e^Y, \quad Y=\ln(X), \quad x_i = e^{y_i}, \quad i =1 , \dots, n,
\ee
Then  $\tau^{(H_q, \beta)}\left(\big[X \big]\right) $ becomes a ratio of Wronskian determinants
\be
\tau^{(H_q, \beta)}\left(\big[X \big]\right) =\gamma_n  \left(\prod_{i=1}^n x_i^{n-1}\right){\det\left(( \phi_n^{({\bf c}, {\bf d}, \beta)})^{(i-1)}(e^{y_j})\right)_{1\leq i,j, \leq n} \over \Delta(e^y)}.
\ee

  
   \subsubsection{Matrix integral representation of $\tau^{(G,\beta)}([X])$: quantum case}
It similarly follows \cite{HR} that
\be
\tau^{(H_q, \beta)}(\big[ X \big])= {\beta^{{1\over 2}n(n-1)} (\prod_{i=1}^n x_i^{n-1})\Delta(\ln(x)) \over (\prod_{i=1}^n i!) 
\Delta(x) } \Zb_{d\mu_{q} }(\ln(X)),
\ee
\bea \text{where} \quad
\Zb_{d\mu_{(q, n)}}(X) &\&=\int_{M \in \Nor^{n\times n}_{\CC_n}}d\mu_{(q,n)}(M) e^{\tr Y M},\cr
\text{and} \quad d\mu_{(q,n)} (M)&\&:= (\Delta({\bf \zeta})^2 \det(A_{H_q, n}(M))
\nonumber
 \eea
 is a conjugation invariant measure on the space of normal matrices  
 \be
M = U Z U^\dag \in \Nor^{n\times n}_{\CC_n}, \quad U \in U(n), \quad Z= \diag(\zeta_1, \dots, \zeta_n)
\ee
with eigenvalues $\zeta_i \in \Cb$ supported on the contour $\CC_n$.

 \bigskip
\noindent 
\small{ {\it Acknowledgements.} 
This work was partially supported by the Natural Sciences and Engineering Research Council of Canada (NSERC) 
and the Fonds de recherche du Qu\'ebec, Nature et technologies (FRQNT).

    \bigskip
    \noindent


\bigskip

  

     \begin{thebibliography}{99}
  
  
  \bibitem{ACEH1} A.~Alexandrov, G. Chapuy, B. Eynard and J. Harnad, 
``Weighted Hurwitz numbers and topological recursion: an overview'', 
{\em J. Math. Phys.} {\bf 59},  081102: 1-20 (2018).

    \bibitem{GH1} M. Guay-Paquet and J. Harnad,
 ``2D Toda $\tau$-functions as combinatorial generating functions'',  
 {\em Lett. Math. Phys.} {\bf 105,} 827-852 (2015).
 
 \bibitem{HO1} J. Harnad and A. Yu. Orlov, ``Hypergeometric $\tau$-functions, Hurwitz numbers and enumeration of paths'',
  {\em Commun. Math. Phys. } {\bf 338}, 267-284 (2015).
  
    \bibitem{H1} J. Harnad,  ``Weighted Hurwitz numbers and hypergeometric $\tau$-functions: an overview'', 
 {\em AMS Proceedings of Symposia in Pure Mathematics} {\bf  93}, 289-333 (2016).
 
  \bibitem{GH2} M. Guay-Paquet and J. Harnad,  ``Generating functions for weighted Hurwitz numbers'', 
  {\em J. Math. Phys.} {\bf  58}, 083503 (2017).
  
  \bibitem{BH} M. Bertola and J. Harnad,  ``Rationally weighted Hurwitz numbers, Meijer $G$-functions
  and matrix integrals'',   {\em J. Math. Phys.}  {\bf 60}, 103504 (2019).
    
   \bibitem{HR} J. Harnad and B. Runov,  ``Matrix model generating function for quantum weighted Hurwitz numbers'',  
   {\em J. Phys.  A} {\bf 53}, 065201 (2020).
      
   \bibitem{Ok} A.~Okounkov, ``Toda equations for Hurwitz numbers'', {\em Math.~Res.~Lett.} {\bf 7}, 447-453 (2000).

\bibitem{OP}  A.~Okounkov and R.~Pandharipande,   ``Gromov-Witten theory, Hurwitz theory and completed cycles,''
  {\em Ann.\ Math.}  {\bf 163}, 517-560 (2006).
    


  \end{thebibliography}
    \end{document}